\documentclass[correspondence]{IEEEtaes}

\usepackage{color,array,amsthm,amsmath,amssymb}
\ifCLASSOPTIONcompsoc
\usepackage[caption=false,font=normalsize,labelfon
t=sf,textfont=sf]{subfig}
\else
\usepackage[caption=false,font=footnotesize]{subfig}
\fi
\usepackage{graphicx}
\usepackage{physics}
\usepackage{myflushend}
\DeclareMathOperator{\diag}{diag}

\jvol{XX}
\jnum{XX}
\jmonth{XXXXX}
\paper{}
\pubyear{2025}
\doiinfo{TAES.2025.Doi Number}

\newcommand*{\tran}{^{\mkern-1.5mu\mathsf{T}}}

\begin{document}

\sptitle{Correspondence}
\title{Transformer Based Multi-Target Bernoulli Tracking for Maritime Radar} 

\author{CADEN SWEENEY}
\author{DU YONG KIM}
\author{BRANKO RISTIC}
\affil{School of Engineering, RMIT University, Melbourne, VIC, Australia} 
\author{BRIAN CHEUNG}
\affil{Defence Science and Technology Group, Edinburgh, SA, Australia}

\receiveddate{Manuscript received XXXXX 00, 0000; revised XXXXX 00, 0000; accepted XXXXX 00, 0000.\\}

\corresp{}

\authoraddress{Authors’ addresses: Caden Sweeney, Du Yong Kim, and Branko Ristic are with the School of Engineering, RMIT University, Melbourne, VIC 3000, Australia, E-mail: (s4032741@student.rmit.edu.au; duyong.kim@rmit.edu.au; branko.ristic@rmit.edu.au); Brian Cheung is with the Defence Science and Technology Group, Edinburgh, SA 5111, Australia, E-mail: (brian.cheung@defence.gov.au). (Corresponding author: Caden Sweeney.)}

\markboth{CORRESPONDENCE}{}
\maketitle

\begin{abstract}
Multi-target tracking in the maritime domain is a challenging problem due to the non-Gaussian and fluctuating characteristics of sea clutter. This article investigates the use of machine learning (ML) to the detection and tracking of low SIR targets in the maritime domain. The proposed method uses a transformer to extract point measurements from range-azimuth maps, before clustering and tracking using the Labelled mulit-Bernoulli (LMB) filter. A measurement driven birth density design based on the transformer attention maps is also developed. The error performance of the transformer based approach is presented and compared with a constant false alarm rate (CFAR) detection technique. The LMB filter is run in two scenarios, an ideal birth approach, and the measurement driven birth approach. Experiments indicate that the transformer based method has superior performance to the CFAR approach for all target scenarios discussed.
\end{abstract}

\correspauthor%
\section{INTRODUCTION}
Detection and tracking of maritime targets in sea clutter is difficult due to several factors. Clutter returns are non-Gaussian and fluctuate temporally and spatially due to the dynamics of the sea surface \cite{7738356}. 

In multi-target tracking (MTT), false alarms and missed detections are unavoidable using methods that achieve CFAR \cite{https://doi.org/10.1049/iet-rsn.2018.5064}. Even with detectors that achieve this constraint, spurious detections that exist over a number of scans could spawn false tracks \cite{doi:10.1049/PBRA025E}. In an environment with spiky and fluctuating clutter, false tracks occur as the threshold at which targets are declared must be sufficiently low to detect targets with weak signal returns.

To reduce the number of false alarms that the tracking algorithm must deal with, techniques that adaptively set the detection threshold are used. Typically these adaptive threshold algorithms estimate the interference level and adjust the threshold to maintain a CFAR. A widely used approach known as cell averaging CFAR (CA-CFAR) \cite{Richards2022}, computes a detection threshold  for a cell based on the estimated mean interference level of adjacent cells. CA-CFAR has been shown to perform optimally when clutter amplitude remains constant between cells, however this approach can cause high false alarms when applied in environments with fluctuating clutter amplitude \cite{9114828}. 

The authors in \cite{9624631}, suggest that when using CFAR detection strategies in a tracking scenario, complexity and computational requirements can increase especially with low signal to interference (SIR) targets. The analysis completed in \cite{https://doi.org/10.1049/iet-rsn.2018.5064} supports these conclusions suggesting that tracking in sea-clutter reduces the performance of tracking algorithms due to the increase in false alarms.

With recent advances in ML and artificial intelligence (AI), there have been several attempts to include these technologies in the context of target detection and tracking. The work in \cite{Peters2021} details an approach using separate detection and tracking schemes for low SIR targets. The use of an ML target detector with a Bernoulli tracker in a hybrid approach was explored in \cite{Sweeney2024}. Methods proposed in \cite{Wang2019} use ML to complete binary classification of target existence which show increase in performance over traditional threshold methods, however do not consider low SIR or complex cluttered environments. Multi-target tracking algorithms that use ML transformers are shown in \cite{Pinto2021}, \cite{Pinto2023}, \cite{Zhao2022}. Learning based methods have shown increase in tracking accuracy over classical methods, when applied to the target tracking problem.

This paper proposes a multi target tracking algorithm that uses a transformer network to provide detections in fluctuating clutter, prior to using the LMB for tracking. The novel contributions of this paper are:

\begin{enumerate}
    \item A low SIR multi-target detection Transformer algorithm that uses a LMB filter to complete tracking on the detections.
    \item A measurement driven birth density design for the LMB based on transformer attention map sampling.
\end{enumerate}

The paper is organised as follows. Section \ref{problem} describes data generation and motion modelling used for this study. Section \ref{tracking} outlines the design of the detection and tracking algorithm. Section \ref{results} describes parameters used, results and a comparison of the newly developed method while Section \ref{conclusion} concludes the study.

\section{BACKGROUND} \label{problem}
\subsection{Radar Data and Parameters}
\begin{figure}
\centerline{\includegraphics[width=18.5pc]{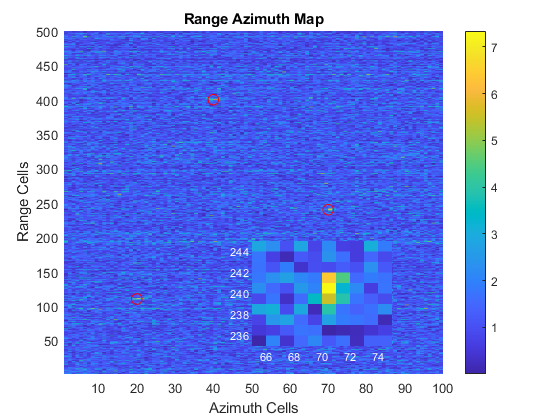}}
\caption{Range-azimuth intensity map. Ground truth targets are highlighted by the red circles with returns of 8dB.}
\label{fig:Range-azimuth}
\end{figure}

Radar data available for detection and tracking are represented by range-azimuth magnitude maps from an airborne scanning radar. At each discrete time step \(k\) a new range-azimuth map \(z_k\) is received from the radar signal processing unit. Each cell of the range-azimuth map is assumed to contain the modulus of the complex returned radar signal. A K-distribution is used to model magnitude fluctuation in \(z_k\), and spatial correlation is modelled using a Gaussian function. If present, the target may spread over several range-azimuth cells which is typically modelled using a Gaussian point spread function. The power contributed by each target is calculated by SIR \(= 10 log_{10}(\chi/\mu)\), where \(\chi\) is the mean amplitude of the target return and \(\mu\) is the mean amplitude of clutter. For this study targets that have returns of \(\chi = [3\text{dB},\ 5\text{dB},\ 8\text{dB} ]\) are considered. A range-azimuth map, with targets classified by 8dB SIR is shown in Fig. \ref{fig:Range-azimuth}. The red circles indicate the locations of each of the targets present in the surveillance volume. The close up portion of the range-azimuth map in Fig. \ref{fig:Range-azimuth} indicates a target with power that spreads into adjacent cells.

\subsection{Target Motion Models}
The target is modelled using a discrete-time motion model, with sampling period \(T\). The target state is defined at time index \(k\), by the state vector \(\mathbf{x}_k = [x_k\:, y_k\:, \Dot{x}_k\:, \Dot{y}_k\:]^{\tran}\), where \((x_k, y_k)\) is the position and \((\Dot{x}_k, \Dot{y}_k)\) is the velocity of the target in Cartesian coordinates. The target state changes according to the constant velocity model \cite{10.1002} with a linear Gaussian transition density given by:
\begin{equation}
    \pi_{k|k-1}(\mathbf{x}|\mathbf{x}') = \mathcal{N}(\mathbf{x};\mathbf{F}_{k-1}\mathbf{x}',\mathbf{Q}_{k-1}),
\end{equation}
where \(\mathcal{N}(\mathbf{x;m,P})\) is a Gaussian probability density function (PDF) evaluated at $\mathbf{x}$ with mean $\mathbf{m}$ and covariance $\mathbf{P}$. The matrices $\mathbf{F}$ and $\mathbf{Q}$ are the transition and process noise covariance matrices respectively. They are defined as:
\begin{equation}
    \mathbf{F} = \begin{bmatrix}
        \mathbf{\Phi} & 0 \\
        0 & \mathbf{\Phi} \\
    \end{bmatrix}, \mathbf{\Phi} = \begin{bmatrix}
        1 & T \\
        0 & 1
    \end{bmatrix},
\end{equation}
\begin{equation}
    \mathbf{Q} = \begin{bmatrix}
        \mathbf{\Psi} & 0 \\
        0 & \mathbf{\Psi} \\
    \end{bmatrix}, \mathbf{\Psi} = \begin{bmatrix}
        T^3/3 & T^2/2 \\
        T^2/2 & T
    \end{bmatrix}q,
\end{equation}
where $q$ is the process noise intensity in the spatial domain.

\section{TRANSFORMER BASED MULTI-TARGET TRACKING} \label{tracking}

\subsection{Transformer Detection} 

\subsubsection{Transformer Architecture} 
Machine learning approaches have shown improvements in detection accuracy and speed when applied to the radar target tracking problem \cite{Sweeney2024}. The transformer architecture used for the work completed in this study is based on the DEtection TRansformer (DETR) \cite{Carion2020}. We adopt an approach similar to \cite{Sweeney2024}, where the attention block in the transformer is used for target detections. 

\textit{Attention}:
Attention is described as a mapping of a query and key-value pair to an output, where the output is then weighted by compatibility between a query and its corresponding key \cite{Vaswani2017}.  The input encoded sequence is mapped linearly into queries (Q), keys (K) and values (V). These sequences are calculated as follows:
\begin{equation}
    Q = W_Q A,
    K = W_K A,
    V = W_V A,  
\end{equation}
where $W_Q, W_K$ and $W_V$ are learnable matrices of the self attention mechanism and $A$ is the input. Attention weights are computed by taking a softmax of the dot products of queries with all keys. The matrix of output attention is computed by:
\begin{equation}
    Attention(Q,K,V) = softmax(\frac{QK^T}{\sqrt{d_k}})\cdot V
\end{equation}
where \(d_k\) is the dimension of the keys. The dimension of the keys is a fixed value and for this study we use \(d_k = 256\). The model then employs multi-head attention, which enables the model to attend jointly to information from different positions, while using the entire image space as context \cite{Vaswani2017}. This is computed by concatenating all of the attention heads \(h = [1, \dots,8]\), where each of the heads are
$head_h = Attention(QW_h^Q, KW_h^K, VW_h^V)$. 

The multi-head attention layer is used in multiple ways throughout the model. There is a form of cross attention between encoder and decoder structures, where queries from the previous decoder layer, and keys and values come from the encoder layer. This enables the decoder to attend to all positions in the input sequence. It is also used in the encoder and decoder structures enabling attention to all positions in the previous layer \cite{Vaswani2017}.
\subsubsection{Attention as detections}\label{attention_dets}
\begin{figure}[ht]
\centering
\includegraphics[width=22pc]{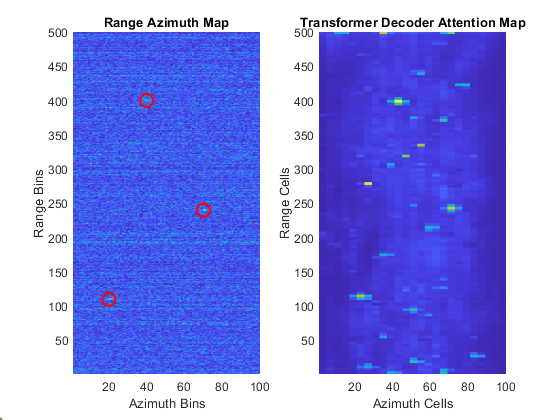}
\subfloat[]{\hspace{.5\linewidth}}
\subfloat[]{\hspace{.5\linewidth}}
\caption{(a) Input range-azimuth map to the transformer. Targets highlighted by red circles. (b) Multi-target attention map of the last transformer decoder layer showing a transformed version of the input.}
\label{fig:extracted}
\end{figure}

In the multi-target scenario, the transformer is trained to detect if targets exist, as well as their spatial locations within the surveillance volume. Similarly to the work in \cite{Sweeney2024}, the attention map from the last decoder layer is used as a transformed representation of all possible targets within the range-azimuth map. The transformer architecture in \cite{Carion2020} allows for a varying number of maximum objects to be detected per image. This is referred to as the number of object queries \(o_{1:n}\) used in the model, where \(n\) must be set significantly higher than the true number of maximum objects through the observation scenario. The transformer creates an attention map for each of the object queries. We formulate the multi-target attention map by summing each of the \(n\) attention maps. Fig. \ref{fig:extracted}(b) indicates the multi-target attention map that corresponds to the input range-azimuth map in Fig. \ref{fig:extracted}(a). Attention maps are then processed to produce point measurements to be input to the tracker.

\subsubsection{Transformer Measurement model}
Due to the backbone in the transformer, the input range-azimuth maps are down sampled to produce the features that are used in the encoder and decoder. Before computing the spatial locations of targets, nearest interpolation is used to up-sample the multi-target attention map to the same coordinate space as the initial range-azimuth map. 

Measurements are obtained from the multi-target attention map by applying a threshold. This threshold, denoted \(t\), must be set so that all potential targets are found. False detections may be included in the measurement set but will be filtered out by the Bernoulli tracker. Measurements from single targets within the multi-target attention map could span multiple cells. Clustering is required to convert target signatures from the attention map into point measurements for tracking. For this study we use density-based spatial clustering (DBSCAN) \cite{10.5555/3001460.3001507}. DBSCAN is robust against outliers and has flexibility when there is a unknown number of clusters. The parameters for the DBSCAN clustering algorithm are the minimum number of points to form a cluster, \(\kappa\) and the maximum distance between points in a cluster \(\epsilon\). 

A set of measurements \(\mathbf{Z}_k = \{\mathbf{z_1, \dots, \mathbf{z_{m}}}\}\) that contains equivalent point measurements corresponding to the centres of clusters is obtained from the multi-target attention map at scan \textit{k}. There is a random number of point measurements, \textbf{m} at each scan.

The probability of detection is denoted as \(\textit{p}_D\), the spatial distribution of false points over the measurement space is \(c(\mathbf{z})\) and the number of false points per scan is assumed to be Poisson distributed with mean \(\lambda\). When using a set of measurements, and assuming that \(\textit{p}_D\) is state independent, the Bernoulli filter can be updated with the measurement likelihood function \(g_k(\mathbf{z}|\mathbf{x})\), where \(\mathbf{z} \in \mathbf{z}_k\). This assumes that the measurement originates from a target. 

The measurement likelihood function assumes a Gaussian distribution:
\begin{equation}
     g_{k}(\textbf{z}|\textbf{x}) = \mathcal{N}(\textbf{w};\textbf{H}\textbf{x},\textbf{R}),
\end{equation}
where 
\begin{equation}
    \mathbf{H} = \begin{bmatrix}
        1 & 0 & 0 & 0 \\
        0 & 0 & 1 & 0
    \end{bmatrix}
\end{equation}
and \(\mathbf{R}\) is the covariance matrix associated with the measurement set.
\begin{equation}
    \mathbf{R}_k = \begin{bmatrix}
        \sigma_A^{2} & 0 \\
        0 & \sigma_R^{2}\\
    \end{bmatrix}
\end{equation}
where \(\sigma_A^{2}\) and \(\sigma_R^{2}\) are the standard deviation of measurement noise in azimuth and range respectively.

\subsection{Labelled Multi-Bernoulli Filter}\label{tracker}
The Labelled Multi-Bernoulli filter is implemented according to \cite{6814305}, as an approximation of the full multi-target \(\delta-\)GLMB posterior recursion. A Gaussian mixture representation is also selected for the LMB filter in this study. There are three distinct steps to achieve target tracking using the LMB filter which include prediction, grouping and parallel updates. We refer the reader to \cite{6814305} for the target grouping implementation. The targets throughout the scenario in this study are well separated which will lead to distinct target groups that contain only a single target.

\subsubsection{Prediction} \label{LMB_prediction}
The posterior of the LMB filter is an LMB distribution with state space \(\mathbb{X}\) and label space \(\mathbb{L}\). It is characterised by the following parameters

\begin{equation}
    \pi = \{(r^{(\ell)}, p^{(\ell)}\}_{\ell \in \mathbb{L}}\ ,
\end{equation}
where \(r\) is the probability of existence and \(p\) is the target state density. The prediction until the time of the next measurement follows an LMB distribution with state space and label space \(\mathbb{L}_+ = \mathbb{L} \ \cup \mathbb{B}\) given by 

\begin{equation}
    \pi_+ = \{(r_{+,S}^{(\ell)},\ p_{+,S}^{(\ell)}\}_{\ell \in \mathbb{L}} \ \cup\ \{(r_{B}^{(\ell)},\ p_{B}^{(\ell)}\}_{\ell \in \mathbb{B}} \ ,
\end{equation}
where the first component represents the surviving Bernoulli components from the previous time step, and the second component represents the labelled multi-Bernoulli birth distribution that is specified \textit{a priori}. Surviving components inherit the same label from the previous time step but are updated using the survival probability and transition density. The birth components are initialised with new unique labels \(\ell \in \mathbb{B}\).

\subsubsection{Update}
In order to use the measurement set \(\mathbf{Z}_k\) from the transformer at the current time step for updates, the predicted LMB density from Section \ref{tracker}.\ref{LMB_prediction} must be expressed as a \(\delta\textit{-}\text{GLMB}\) density. The predicted \(\delta\textit{-}\text{GLMB}\) is given by

\begin{equation}
    \pi_+(\widetilde{\mathbf{X}}_{+}) =\Delta(\widetilde{\mathbf{X}}_{+})\sum_{I^+ \in \mathcal{F(\mathbb{L}_+)}} w_{+}^{(I_+)} \delta_{I_+}(\mathcal{L}(\widetilde{\mathbf{X}}_{+}))[p_+]^{\widetilde{\mathbf{X}}_{+}}
\end{equation}
with 

\begin{equation}
    w_{+}^{(I_+)} = \prod_{\ell \in \mathbb{L}_+} (1-r_+^{(\ell)}) \prod_{\ell' \in I_+} \frac{1_{\mathbb{L}_+(\ell') r_+^{(\ell')}}}{1-r_+^{(\ell')}}
\end{equation}
where \(\widetilde{\mathbf{X}}_{+}\) is the multi-target state.

The \(\delta\textit{-}\text{GLMB}\) update for each of the groups is given by 
 \begin{equation}
 \begin{split}
     \pi_+(\widetilde{\mathbf{X}}_{+}|\mathbf{Z}_k) =\Delta(\widetilde{\mathbf{X}}_{+}) \sum_{(I_+, \theta) \in \mathcal{F}(\mathbb{L}_+)\times \Theta_{I_+}} w^{(I_+, \theta)}(\mathbf{Z}_k) \\
     \times \delta_{I_+}(\mathcal{L}(\widetilde{\mathbf{X}}))\big[p^{\theta}(\cdot|\mathbf{Z}_k)\big]^{\widetilde{\mathbf{X}}}
 \end{split}
 \end{equation}
 where \(\Theta_{I_+}\) represents the mappings \(\theta : I_+ \rightarrow \{0,\dots,|Z\}\), such that \(\theta(\iota) = \theta(\iota') > 0\) implies \(\iota = \iota'\) and 

 \begin{equation}
     w^{(I_{+},\theta)}(\mathbf{Z}_k) \propto w_{+}^{(I_+)} \big[\eta^{(\theta)}_{\mathbf{Z}_k}\big]^{I_+}
 \end{equation}

 \begin{equation}
     p^{(\theta)}(x, \ell|\mathbf{Z}_k) = \frac{p_{+}(x,\ell) \psi_{\mathbf{Z}_k}(x, \ell;\theta)}{\eta_{\mathbf{Z}_k}^{(\theta)}(\ell)},
 \end{equation}

 \begin{equation}
     \eta_{\mathbf{Z}_k}^{(\theta)}(\ell) = \langle p_{+}(x,\ell),\psi_{\mathbf{Z}_k}(\cdot, \ell;\theta)\rangle,
 \end{equation}

 \begin{equation}
     \psi_{\mathbf{Z}_k}(x,\ell;\theta) = \begin{cases}
         \frac{p_D(x,\ell)g(Z_{k,\theta(\ell)}|x,\ell)}{\kappa(Z_{k,\theta(\ell)})}, & \text{if} \ \theta(\ell)>0 \\
         q_{D,G}(x,\ell), & \text{if} \ \theta(\ell) = 0
     \end{cases}
 \end{equation}
 where \(p_D\) is the probability of detection, \(g(Z_k|x,\ell)\) is the single target likelihood and \(q_{D,G}(x,\ell) = 1-p_{D}(x,\ell)p_G\).
 After applying the update equations to each of the tracks the \(\delta\textit{-}\text{GLMB}\) form is converted back to the LMB representation

 \begin{equation}
     \mathbf{\pi}(\cdot|\mathbf{Z}_k) \approx \tilde{\mathbf{\pi}}(\cdot|\mathbf{Z}_k) = \bigl\{(r^{(\ell)},p^{(\ell)})\bigr\}_{\ell \in \mathbb{L}_{+}}
 \end{equation}
 where the multi-target posterior is given by the LMB approximation 

\begin{equation}
     \mathbf{\pi}(\cdot|\mathbf{Z}_k) \approx \tilde{\mathbf{\pi}}(\cdot|\mathbf{Z}_k )= \bigl\{(r^{(\ell)},p^{(\ell)})\bigr\}_{\ell \in \mathbb{L}_+}.
\end{equation}

Once the multi-target posterior has been computed, tracks are declared if the existence probability is above a certain threshold. The posterior density is also used to generate the LMB birth density at the next time step.

\subsection{Measurement Driven Birth Density} \label{Birth}
The implementation of the LMB filter in Section \ref{tracker} relies on knowledge about target birth densities. This requires background knowledge of where targets may spawn, such that birth densities are created with low spatial uncertainties.

To adaptively form the multi-Bernoulli birth density, the transformer attention maps in Section \ref{attention_dets} are sampled to create potential birth locations of tracks. The multi-Bernoulli birth density \(\pi_{B,k}\) at the current scan \(k\) is dependant on the set of measurements \(\mathbf{Z}_{k-1}\) sampled from the multi-target attention map at the previous scan \(k-1\). The spatial density of the multi-Bernoulli birth density is formed from measurements that are not associated to existing tracks. The Euclidean distance between each of the measurements and confirmed tracks is calculated as

\begin{equation} \label{birth_dist}
    d_{(\mathbf{z},\hat{\mathbf{X}})}= \| \mathbf{z} - \hat{\mathbf{x}}\| 
\end{equation}
where \(\mathbf{z} \in \mathbf{Z}_{k-1}\) is the measurement and \(\hat{\mathbf{x}}\) is the position of the target in the confirmed track. If the distance from (\ref{birth_dist}) exceeds a threshold the measurement is included in the spatial density. The threshold used in this study is \(t_d =  10\) cells.

The birth probability is calculated based on the confidence of each detection from the transformer. This ensures that the most confident detections from the transformer are able to spawn tracks more quickly, while reducing the possibility of false tracks from low confidence predictions. The birth probabilities are obtained by

\begin{equation}
    r_{B,k}^{(\ell)}(\mathbf{z}) = r_{B, max}\ \times \hat{p}_{\mathbf{w}}
\end{equation}
where \(\mathbf{z} \in \mathbf{Z}_{k-1}\) is the measurement from the previous time step if it is determined to be a suitable birth location, \(r_{B, max}\) is the maximum birth probability and \(\hat{p}_w\) is the normalised confidence value or prediction score corresponding to measurement \(\mathbf{z}\). The weights and co-variances of the birth density are fixed and set once a new birth component has been initialised.
\section{NUMERICAL RESULTS} \label{results}

\subsection{Experimental Setup}
The performance of tracking algorithms is measured using the mean optimal sub pattern assignment (OSPA) metric \cite{4567674}. This metric penalises the error in cardinality and  position, however does not use velocity or amplitude in the assessment. There are two parameters required for the OSPA metric: order \(\rho\) and cutoff \(c\). For the simulations completed in this study we set \(\rho = 1\). The parameter \(c\) indicates the penalty assigned to the cardinality error in terms of cells, for this study \(c = 100\). The OSPA distance is averaged over 100 Monte Carlo runs. The multi-target sequence involves 50 range-azimuth maps, with targets injected at range-azimuth locations (20,110), (40,400) and (70,240). The targets each have non-zero velocities in the range and azimuth directions but remain well separated. All target amplitudes fluctuate between scans due to the nature of clutter, however their mean returned power remains constant.

The CA-CFAR detector used in this study for comparison has the following parameters. The number of training cells is \(N_{T_{R}} = N_{T_{A}} = 7\), and the number of guard cells is \(N_{G_{R}} = 3\ \text{and}\ N_{G_{A}} = 6 \) where \(R\)
and \(A\) are the cells in range and azimuth. The probability of false alarm is set to \(P_{FA} = 0.001\) and the requested detection threshold is set to detect targets at 3dB.

For this study the detections from the multi-target attention map are declared if the pixel value exceeds a threshold of \(t = 0.1\). Measurements from the transformer detector and the CA-CFAR detector are subject to clustering using DBSCAN \cite{10.5555/3001460.3001507}. This is required as detections may spread in the range and azimuth directions. For the transformer, detections from the attention map are clustered using \(\kappa = 5\) and \(\epsilon = 1\). For the CA-CFAR detections are clustered using \(\kappa = 2\) and \(\epsilon = 4\). These parameters were tuned to maximise the performance of both detection methods.

The CA-CFAR detector output and the transformer detector output after clustering are processed by the same LMB filter to produce output tracks. The parameters used for the LMB filter described in Section \ref{tracking}\ref{tracker} are as follows. The survival probability of all targets is \(p_{S,k} = 0.99\). 

The filter is run using two different birth models, an ideal model where the filter knows the exact starting locations of each of the targets (referred to as ideal birth) and the measurement driven model discussed in Section \ref{tracking}\ref{Birth}. For the ideal birth (IB) model, the density is created as a multi-Bernoulli RFS with the density \(\pi_B = \{r_{B}^{(i)},p_{B}^{(i)}\}_{i=1}^{3}\) where \(r_B^{(1)} = r_B^{(2)} = r_B^{(3)} = 0.03\), \(p_{B}^{(i)}(x) = \mathcal{N}(x;m_{B}^{(i)},P_B), m_{B}^{(1)} = [20,0,110,0],m_{B}^{(2)} = [40,0,400,0],m_{B}^{(3)} = [70,0,240,0], P_B = \diag(100,100,100,100)\). The measurement driven birth (MDB) density, follows a similar density with \(r_{B,max} = 0.04\) and \(m_{B}^{(i)} = [\mathbf{z}_{k-1,A}^{(i)},0,\mathbf{z}_{k-1,R}^{(i)},0]\) where \(\mathbf{z}_{k-1}\) is the measurement \(i\) from the previous scan that creates a component of the spatial density. The covariance of the measurement driven density is the same as the ideal birth model. 

The covariance of observation noise is \(\sigma_A = \sigma_R\ = \sqrt{2.5}\), the process noise intensity is \(q = 0.1\) and the discrete sampling time is \(T=1\). The probability of detection is \(p_D = 0.98 \) and we assume the average rate of clutter per scan to be \(\lambda_c = 30.\)

\subsection{Transformer Training}
\begin{figure}[ht]
\subfloat[]{\label{subfig:a}\includegraphics[width=18.5pc]{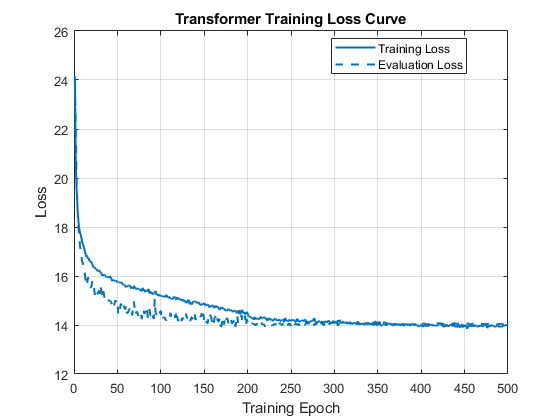}}\\
\subfloat[]{\label{subfig:b}\includegraphics[width=18.5pc]{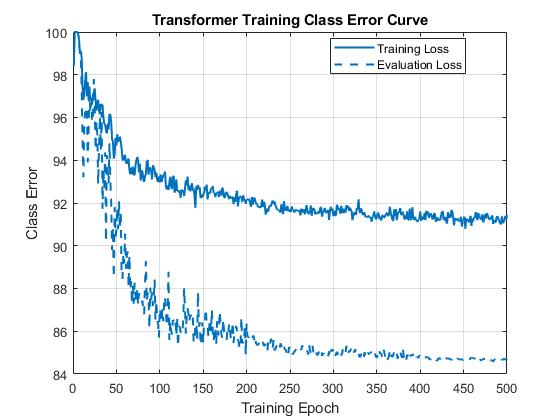}}\\
\subfloat[]{\label{subfig:c}\includegraphics[width=18.5pc]{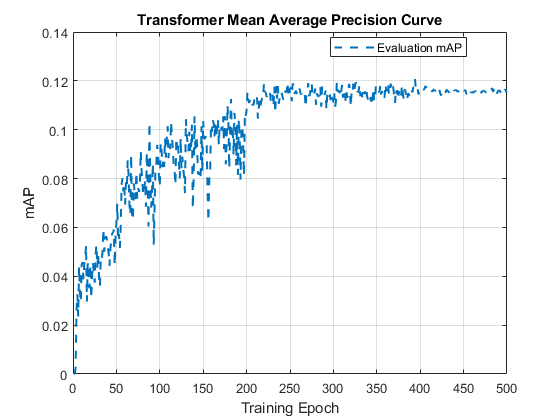}}
\caption{(a) Transformer loss for training and test runs over 500 epochs (b) Transformer class error for training and test runs over 500 epochs (c) Transformer mAP for training runs only over 500 epochs.}
\label{fig:lossCurves}

\end{figure}

The transformer used for this study is trained in a multi-class classification scenario. The classes are split by SIR, where there is a class for no target, 3dB target, 5dB target and 8dB target. The dataset was created using trajectories of targets starting at random locations within the surveillance volume. Each class contains 75 unique trajectories of 50 scans each which creates 3750 images per class, where the total dataset (15000 images) is split \(80\%-20\%\) for train and validation respectively. The transformer is trained from scratch for 500 epochs on a single Quadro RTX 5000 GPU. The training loss curves are shown in Figure \ref{fig:lossCurves}. The train and validation curves are shown for each of the metrics. The loss, class error and mean average precision (mAP) all converge to suitable levels for this application after 500 epochs. The results in the following section also support that training has been completed to a sufficient level.

\subsection{Experimental Results}
\begin{figure}
\centerline{\includegraphics[width=18.5pc]{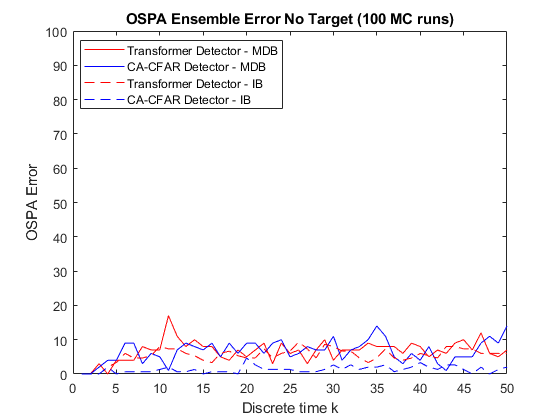}}
\caption{OSPA ensemble average for transformer based and CFAR based algorithms with no targets}
\label{fig:ospa-noTarg}
\end{figure}

The proposed method was evaluated with various experiments, each setup with different target scenarios. The case with no targets present is shown in Fig. \ref{fig:ospa-noTarg}. This evaluates whether the detection and tracking framework creates a high number of false tracks. Both of the algorithms have similar false track rates, although in the case with ideal birth the CA-CFAR detector performs better. When both algorithms use measurement driven birth, the false track rate is approximately equal.

\begin{figure}
\centerline{\includegraphics[width=18.5pc]{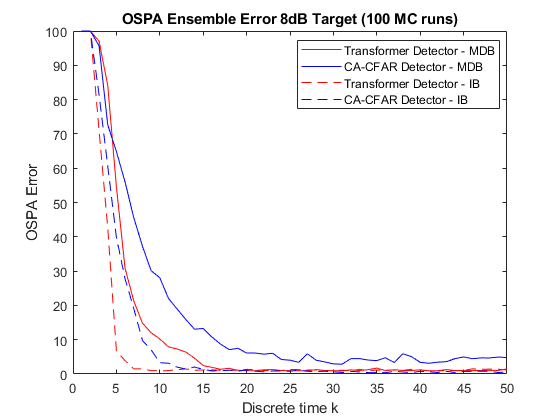}}
\caption{OSPA ensemble average for transformer based and CFAR based algorithms with 8dB target}
\label{fig:ospa-8dB}
\end{figure}

The algorithms are evaluated with targets at 3dB, 5dB and 8dB SIR to simulate performance over varying situations. The Transformer detector outperforms the CA-CFAR detector in all of the scenarios. The error curve in Fig. \ref{fig:ospa-8dB} indicates the tracks are declared quickly in both the CA-CFAR and transformer detector models with ideal birth. When using the measurement driven birth the transformer based method has lower steady state error and declares tracks more quickly indicating better performance than the CA-CFAR detection model.

\begin{figure}
\centerline{\includegraphics[width=18.5pc]{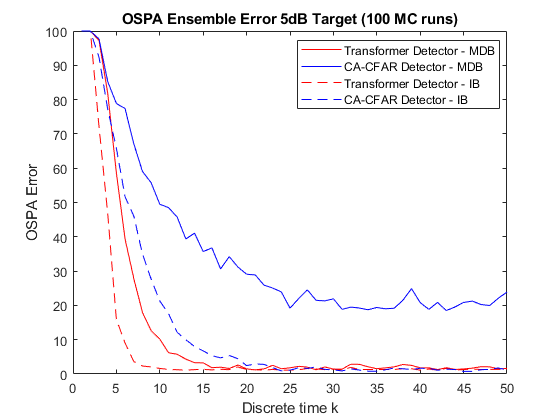}}
\caption{OSPA ensemble average for transformer based and CFAR based algorithms with 5dB target}
\label{fig:ospa-5dB}
\end{figure}

\begin{figure}
\centerline{\includegraphics[width=18.5pc]{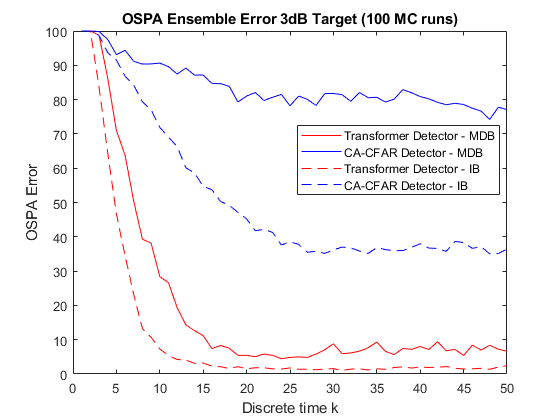}}
\caption{OSPA ensemble average for transformer based and CFAR based algorithms with 3dB target}
\label{fig:ospa-3dB}
\end{figure}

In the case for 5dB targets, Fig. \ref{fig:ospa-5dB} shows that in ideal birth conditions, and measurement driven conditions the transformer approach is able to declare tracks more quickly and produce lower steady state error than the CA-CFAR approach under ideal birth conditions. This observation is also true for targets with 3dB SIR. The CA-CFAR detection rarely detects all of the targets as seen by the high steady state error in comparison to the transformer detector. Even in ideal birth conditions Fig. \ref{fig:ospa-3dB} indicates that the CA-CFAR detector is not able to detect the targets. The transformer is still able to detect and track targets with low error even at low SIR. The error for the 3dB case is higher than 5dB and 8dB, although this is expected as the target returns have characteristics that appear similar to clutter which creates a complex environment for detecting targets.

\begin{figure}
\subfloat[]{\label{subfig:a}\includegraphics[width=18.5pc]{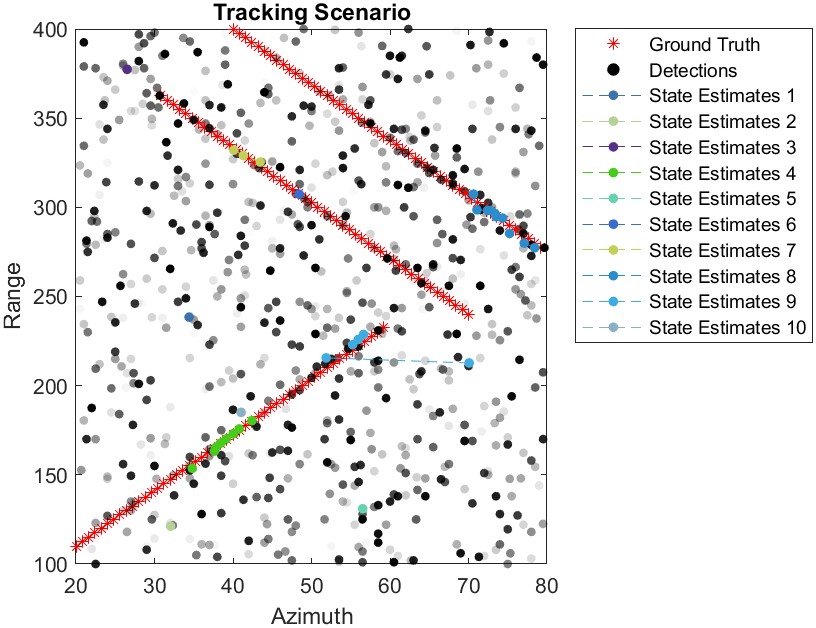}}\\
\subfloat[]{\label{subfig:b}\includegraphics[width=18.5pc]{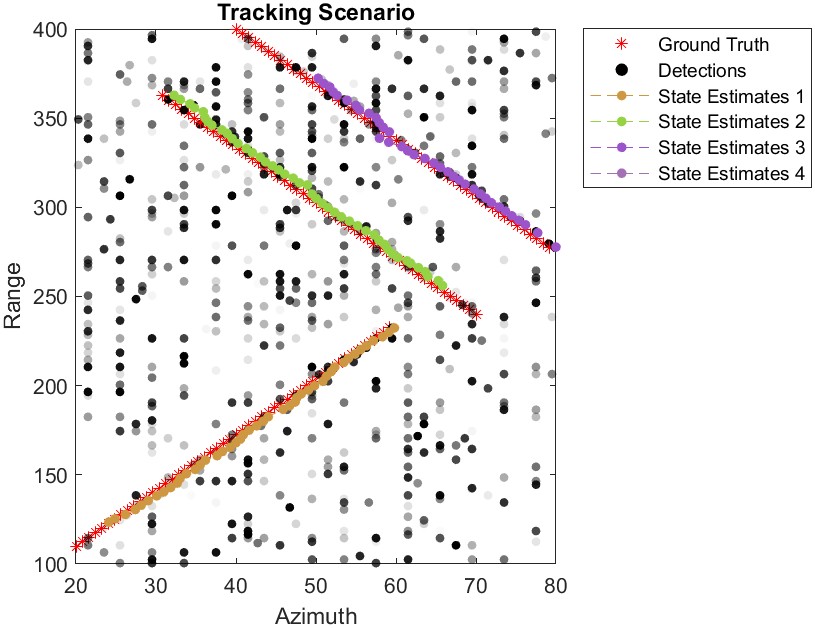}}\\
\caption{(a) Tracking scenario for CA-CFAR detector with tracker state estimates (b) Tracking scenario for transformer detector with tracker state estimates. The ground truth trajectories for 3dB targets are in red. Detections in black are scaled by time where older detections are more transparent}
\label{fig:tracks}
\end{figure}

When the transformer detector and LMB filter is used with the confidence based birth density design, it is able to outperform the CFAR detector and LMB filter when the tracker has prior knowledge about the target birth locations. This indicates that the proposed method has far superior performance at both high and low SIR values. This is illustrated by the single run results plotted in Fig. \ref{fig:tracks}. These plots show the tracking results when using the measurement driven birth density for the CFAR detector and the transformer detector with a 3dB target. We can observe that the transformer-based method successfully tracks all three targets, while the CFAR method struggles to establish confident tracks.

\subsubsection{Computation Time}
Computation time for the detection and filtering stages of each algorithm is measured and averaged across multiple MC runs. Table \ref{tab:comp_time}. shows the computation times. The scan column represents the time taken for each detector to process a single range-azimuth map. The filtering demonstrates the time taken for the LMB filter to operate on the measurements. The total column is the total time taken to process the 50 range-azimuth maps and complete filtering.

\begin{table}
    \centering
    \caption{Computation Times for CA-CFAR and Transformer based approaches}
    \label{tab:comp_time}
    \begin{tabular}{ccccl}\hline
         Algorithm &  Scan (s) & Filtering(s) &Total(s)\\\hline
         CA-CFAR + LMB&  0.2655&  49.2770 & 62.5520\\
         Transformer + LMB&  0.6639&  48.7187 & 81.9137\\
    \end{tabular}
\end{table}
The results show that the CA-CFAR detector is able to process the range-azimuth maps more quickly, but incurs a slightly larger cost to filtering. This is due to a larger number of detections given to the LMB than the transformer. The transformer based detector operates slower over the range azimuth maps, although produces less detections which are more accurate. The transformer produces less false detections and achieves a higher probability of detection.

The time taken by the transformer could be reduced by creating batches of range-azimuth maps for joint detection. Further work could involve reducing the size and number of parameters in the transformer model. Whilst reducing the computation time of the transformer is critical, it is left as future work. 

\section{CONCLUSION} \label{conclusion}
This paper has proposed a multi-target Bernoulli tracking algorithm that incorporates an ML transformer to detect  targets in maritime radar data. The algorithm developed aims to overcome the challenges associated with the detection of multiple targets in spiky sea clutter. Performance of the newly developed method was evaluated in different scenarios and compared to state-of-the-art (SOTA) methods. The transformer based approach had superior performance in all of the scenarios. A measurement driven birth model that used prediction scores from the transformer was also proposed and evaluated. 

Several directions of future research are envisaged: handling closely spaced targets; a comparison with track-before-detect methods; exploration of transformer-based data association and the reduction of transformer computation time.

\section*{Acknowledgement}
This research is supported by the Commonwealth of Australia as represented by the Defence Science and Technology Group of the Department of Defence

{
\bibliography{multitarg}
}
\bibliographystyle{IEEEtran}

\end{document}